\documentstyle[preprint,eqsecnum,prd,aps]{revtex}
\begin{document} 
\renewcommand{\thefootnote}{\fnsymbol{footnote}}
\setcounter{equation}{0}
\newcommand{\beq}{\begin{equation}}
\newcommand{\eeq}{\end{equation}}
\newcommand{\beqa}{\begin{eqnarray}}
\newcommand{\eeqa}{\end{eqnarray}}
\input{epsf}
\pagestyle{plain}

\preprint{
\vbox{
\halign{&##\hfil\cr
	& ANL-HEP-PR-96-7\cr
        & IFT-UFL-96-6 \cr}}
}
\title{Polarized Double Photon Production
in QCD to Order $\alpha_s$}
\author{Claudio Corian\`{o}$^{a,}$
 and L. E. Gordon$^b$}
\address{
$^a$Institute for Fundamental Theory,
Physics Department,\\
 University of Florida at Gainesville, 32611, FL, USA\\
$^b$ High Energy Physics Division, Argonne National Laboratory,
	Argonne, IL 60439, USA }
\maketitle
\begin{abstract} 
We present a complete order $\alpha_s$ analysis of the process $p p\to 
\gamma \gamma + X$ with polarized initial states,
previously studied by us with leading order structure functions.
We include in our calculation new sets of parton densities evolved in
NLO QCD, such as those of Gehrmann and Stirling and of Gl\"{u}ck et al.,
which incorporate the new anomalous dimensions of Mertig and Van Neerven 
in the evolution equation. A detailed phenomenological analysis is also given 
which includes the photon isolation. Our results indicate that the 
asymmetries, although not very large, should be substantial enough at RHIC 
energies to be measurable in future 
planned experiments.  

\end{abstract}
\vspace{0.2in}

\narrowtext
\section{Introduction}
The study of the spin structure of the nucleons is a fascinating subject
which has attracted a lot of attention both from the theoretical and from 
the experimental side in recent years. From a phenomenological viewpoint, 
for instance, the EMC result has called for a re-analysis of the connections 
between the traditional parton model and the QCD description of polarized DIS,
based on leading-power factorization theorems and on applications of the 
Operator Product Expansion.  

There seems to be more agreement now in the literature on the interpretation of 
the EMC and of other more recent results than before, and it seems obvious that
even in a perturbative framework (such as the OPE), subtle issues connected 
to the renormalization of anomalous operators, their scale dependence etc., 
had been overlooked in the past. Sum rules, largely inspired by the naive 
quark model, have also provided a bridge between the theory and the 
experiments. 

It is expected in few years time that the information gathered from DIS 
studies of the polarized parton distributions will be supplemented and
even extended by new results obtained from proton-proton colliders, such as 
the BNL Relativistic Heavy Ion Collider, RHIC. In particular, in these 
experiments a direct gluon coupling will allow direct measurements of the
contributions of the gluons to the spin of the nucleon. 

Recently, new sets of polarized parton distributions have been generated by 
various groups which incorporate the effects of the evolution equation
up to $O(\alpha_s)$. This has been possible thanks to the ground-breaking 
calculation of Mertig and Van Neerven, later verified by Vogelsang \cite{MVN} 
who have derived the $O(\alpha_s)$ radiative corrections to the anomalous 
dimensions of the evolution. 

We have presented a study of the process $p \,\,p 
\to \gamma\,\, \gamma\, +\, X$ to next-to-leading order containing a brief 
phenomenological analysis of our results \cite{corgor}. In this work we are 
going to extend this analysis for the same process by incorporating, for the 
first time, the new NLO sets of structure functions. Here most of our 
discussion, therefore, will be purely phenomenological and we refer to our 
previous study for all the technical details and for a complete presentation 
of the methods involved in it. 

This work is organized as follows. In section II, in order to make 
our discussion self contained, we briefly review the structure of the evolution 
equations for the structure functions to order $\alpha_s$, and discuss in 
general terms the various parametrizations. We then move to a detailed 
phenomenological study of double prompt photon production, and examine
in particular whether it will be possible to get any new information on
the polarized gluon distributions $\Delta G$ from a study of this
process at RHIC. We will also discuss the uncertainties and the scale 
dependence of NLO results, and we make predictions of the large transverse 
momentum ($p_T$) behaviour of the cross section at RHIC.  Our conclusions are 
presented in section IV.

\section{ The NLO Evolution}
Different sets of polarized structure functions to NLO have been proposed 
recently by Gehrmann and Sterling (G-S)\cite{GS} and by Gl\"{u}ck et al. (GRSV)
\cite{GRSV}. For the G-S sets, three different parametrizations are considered,
while for GRSV two possible scenarios have been analyzed. In order to render 
our treatment self contained, we briefly summarize these new results.
We start from the usual framework of a DIS (longitudinal) polarized scattering. 

In the QCD parton model we define 
\beqa
&& q=q^\uparrow + q^\downarrow \nonumber \\
&& \Delta q=q^\uparrow - q^\downarrow
\eeqa
to be the unpolarized and polarized distributions respectively.
$q^\uparrow$ and $q^\downarrow$ are the distributions which describe the 
probability for finding 
quarks with spin parallel or antiparallel to the longitudinally polarized 
nucleon. 
For instance, if we use values for the factorization scale and the 
renormalization 
scale both equal to $Q^2$, we can write down 
the relation between the polarized structure function $g_1$ and the polarized 
quark ($\Delta q(x)$) and gluon distributions ($\Delta G(x)$) by 

\beqa
 g_1(x,Q^2)&=&\int_{0}^1 {d y\over y}\left( 
C_q^S(x/y,\alpha_s(t))\Delta \Sigma(y,t) +
C_q^{NS}(x/y,\alpha_s(t))\Delta q^{NS}(y,t)
\right.\nonumber \\
&+&\left.2 n_f C_g(x/y,\alpha_s(t))\Delta G(y,t)\right),
\nonumber \\
\eeqa
where 
\beqa
 \Delta \Sigma(x,t)&=&\sum_{i=1}^{n_f}\left(\Delta q_i(x,t) +
\Delta \bar{q}_i(x,t)\right)\nonumber \\
\Delta q_{NS}&=&\sum_{i=1}^{n_f}{e_i^2-\langle e^2\rangle\over \
\langle e^2 \rangle}\left(\Delta q_i(x,t) +\Delta \bar{q}_i(x,t)\right).
\eeqa
are respectively the singlet and the non-singlet quark distributions. 
$n_f$ is the number of flavors $(i)$, each of charge $e_i$, 
$\langle e^2\rangle=\sum e_i^2/n_f$ and $t=\log (Q^2/\Lambda^2)$

The singlet part of the Altarelli-Parisi evolution is given by 
\beqa
&& {d\over dt}\Delta \Sigma(x,t)={\alpha_s (t)\over 2 \pi}
\int_x^1{d\,y\over y}\left[ P_{qq}^S(x/y,\alpha_s(t))\Delta\Sigma(y,t) 
+2 n_f P_{qg}(x/y,\alpha_s(t)\Delta G(y,t)\right],\nonumber \\
&&
 {d\over dt}\Delta G(x,t)={\alpha_s (t)\over 2 \pi}
\int_x^1{d\,y\over y}\left[ P_{gq}^S(x/y,\alpha_s(t))\Delta\Sigma(y,t) 
+2 n_f P_{gg}(x/y,\alpha_s(t)\Delta G(y,t)\right].
\eeqa
while the non-singlet part evolves independently 
\beq
 {d\over dt}\Delta q^{NS}(x,t)={\alpha_s (t)\over 2 \pi}
\int_x^1{d\,y\over y} P_{qq}^{NS}(x/y,\alpha_s(t))\Delta q^{NS}(y,t). 
\eeq

At NLO the expansion in power of $\alpha_s$ of the coefficient functions (C) 
and of the splitting functions (P) reads 

\beqa
&& C(x,\alpha_s)=C^{(0)}(x) +{\alpha_s\over 2 \pi}C^{(1)}(x)+ O(\alpha_s^2),
\nonumber \\
&& P(x,\alpha_s)=P^{(0)}(x) +{\alpha_s\over 2 \pi}P^{(1)}(x) +O(\alpha_s^2).
\nonumber \\
\eeqa

The first moment of $g_1(x, Q^2)$ measures the expectation value of 
a combination of 
octet and singlet axial vector currents 
\beq
\int_0^1 g_1(x) d\,x  ={1\over 12}a_3 +{1\over 36} a_8 +{1\over 9}a_0,
\eeq

where 
\beqa
 a_8 &=& 3 F - D\nonumber \\
&=& \Delta u +\Delta\bar{u} +\Delta d 
 +\Delta\bar{d} -2(\Delta s +\Delta\bar{s})=\,\,0.579 \pm 0.025;\nonumber \\
\label{c2}
\eeqa
\beqa
a_3&=&g_A= F + D\nonumber \\
&=& \Delta u +\Delta\bar{u} -\Delta d -\Delta\bar{d}=1.2573 \pm 0.0028,\nonumber 
\\
\label{c1}
\eeqa
\beqa
 a_0&=&\sum_q\Delta_q +\Delta\bar{q}=\Delta q_8 +3(\Delta s +\Delta\bar{s}).
\nonumber \\
\label{c3}
\eeqa
$\Delta q$ denotes the first moment of the corresponding distributions 
$\Delta q(x)$. 
Under suitable assumptions \cite{WA2}, an analysis done in 1983 gives 
for the matrix elements controlling the $\beta$-decay of the spin $1/2$ 
hyperons 
\beq
F= 0.477\pm 0.012\,\,\,\,\,\,\,\,\,\,\,\, D=0.756\pm 0.01.
\eeq

In a more recent analysis (1988) \cite{HSU}
\beq
F=0.46 \pm 0.01\,\,\,\,\,\,\,\,\,\,\,\,\,\,\, D=0.79\pm 0.001, 
\eeq
which implies $a_0=0.06\pm 0.12 \pm 0.17$ if the EMC data 
($\Gamma_1^p[<Q^2>=10.7]=0.128\pm 0.013\pm0.019$) or the SMC data 
($\Gamma_1^p[<Q^2>=10]=0.136\pm 0.011\pm0.011$) for the first moment of $g_1$ 
are used. This value is smaller than the value $a_0=0.188\pm 0.004$ 
expected on the basis of the Ellis-Jaffe sum rule.
The values of $a_3$ and $a_8$ are constraints equations which have been used in 
most of the LO analysis of the polarized structure functions done so far. 
These two constraints are obtained under different assumptions. 
For instance (\ref{c1}) requires $SU(2)_f$ symmetry between 
the matrix elements of the charged and neutral axial currents, while (\ref{c2}) 
requires an $SU(3)_f$ symmetry between matrix elements 
of hyperon decays of charged and neutral weak axial currents. 
This last assumption has been criticized by Lipkin \cite{Lip}. 
He has argued that the $\beta$ decay of hyperons only fix the total helicity of 
the {\em valence} quarks. Therefore eqs.
(\ref{c2}) and (\ref{c1}), in this scenario, should be replaced by 
\beq
 a_8=\Delta u_v +\Delta d_v
\eeq
and 
\beq
a_3= \Delta u_v -\Delta d_v 
\eeq
respectively. 
These two scenarios, the standard scenario and the {\em valence} one, generate
two different sets of parton distributions GRSV1 and GRSV2 in ref.\cite{GRSV}.
The expression of these two parametrizations can be found in the original 
work. Notice that these two scenarios, of course, require different 
assumptions on the contribution to the polarization coming from the {\em sea} 
quarks.

In fact, the two expressions for the first moment of $g_1$, assuming the 
validity of the Bjorken sum rule for $g_A$ (eq.\ref{c1}) and of $SU(3)_f$ 
flavour symmetry are given by, 

\beq
\Gamma_1^p=\left[{1\over 12}(F +D) +{5\over 36}(3 F -D) +
{1\over 3}(\Delta s +\Delta\bar{s})\right]
\left(1-{\alpha_s(Q^2)\over \pi}\right)
\label{g1}
\eeq
for the standard scenario, and 
\beq
\Gamma_1^p=\left[{1\over 12}(F +D) +{5\over 36}(3 F -D) +
{1\over 18}(10\Delta\bar{q} +\Delta s +\Delta\bar{s})\right]
\left(1-{\alpha_s(Q^2)\over \pi}\right)
\label{g2}
\eeq
for the valence scenario. Notice that in the standard scenario (eq. \ref{g1}), 
a negative 
$\Delta s$ has to be required in order to bring down $\Gamma_1^p$ to the 
experimental EMC or SMC result. As we just mentioned, the condition 
$\Delta s=\Delta\bar{s}=0$ is in fact still too high to match the data

In the valence scenario of Lipkin (eq. \ref{g2})
- even with zero strange quark contribution - 
a negative $\Delta\bar{ q}$ is needed in order to obtain the same reduction. 
In general, a finite $\Delta s(Q^2)$ is generated 
by the evolution equations due to the non vanishing of the NLO anomalous 
dimension $\Delta\gamma_{qq}$. This happens for $Q^2>\mu^2$, where $\mu$ 
is the renormalization scale to NLO where the evolution starts.
In GRSV $\mu^2=0.34$ GeV$^2$. 

Let's remark, at this point, that our understanding of higher twist effects 
and renormalon effects at small $Q^2$, which influence the sum rules 
and might affect the evolution, are not yet under control from a theoretical 
perspective. It is estimated, though, that at $Q^2=2$ Gev$^2$, 
for instance, higher twist contributions should be of the order of $10 \%$ 
\cite{Ji}.

It should be pointed out that it is quite common in the literature to find 
different expressions for $\Gamma_1^p$ depending over whether the anomalous 
contribution is included or not in the definition of $a_0$. 
Notice that this amount to a redefinition of $\Delta\Sigma$ of the form 
\beq
\Delta\Sigma\to \Delta\Sigma - {\alpha_s(Q^2)\over 2 \pi}\Delta G
\eeq
which modifies the contribution to $g_1$ which can be attributed 
to the net helicity of 
the quarks by an anomalous gluonic part. 
However, from a 
practical viewpoint, it is convenient to remove it from the LO expression 
of $\Gamma_1$ and let it reappear through the evolution equation (in the 
$\overline{MS}$ scheme), now known 
to order $\alpha_s$, as an additional term in the splitting functions. 
The calculation of the NLO anomalous dimensions of ref. \cite{MVN} are performed 
in the 
$\overline{MS}$ scheme. 

Generically, the ansatz chosen by GRSV for the polarized parton distributions is 
of the form 
\beq
\Delta q(x, Q^2)\sim x^\alpha (1-x)^\beta q(x,Q^2)
\eeq
and a similar form for the gluon content, where the unpolarized 
distributions $q(x,Q^2)$ are taken from \cite{GRV}. The gluon distribution 
is only weakly constrained since, in LO, it doesn't appear in the expression of 
the asymmetries. Various arguments, based on Regge theory or, as in 
ref.\cite{bd}, arguments based on coherence effects, allow to estimate a 
behaviour of the form 
\beq
\left.{\Delta G\over G }\right|_{x\to 0}\sim x.
\eeq
Notice that the extrapolation in $x$ of the various distribution is a 
remarkable source of complexity in spin physics, given the fact that the 
measurements are performed at different values of $Q^2$ at each bin $x$ and an 
intermediate AP evolution to common $Q^2$ is required. 

The G-S distributions in ref. \cite{GS} are generated by global fits of the 
form 
\beq
x\Delta Q=\eta A x^a(1-x)^b (1+\gamma x +\rho \sqrt x)
\eeq
 where three different 
parametrizations are given (here denoted G-S $a$, G-S $b$, G-S $c$). They 
differ in the 
parametrization adopted for the polarized gluon distributions. 
Specifically 

\begin{eqnarray}
{\rm G-S}\;\; a\; &\gamma_G=0  &\;\; \rho_G=0,\nonumber \\
{\rm G-S}\;\; b\; &\gamma_G=-1 &\;\; \rho_G=2,\nonumber \\
{\rm G-S}\;\; c\; &\gamma_G=0  &\;\; \rho_G=-3.
\end{eqnarray}

We refer the reader to the original work \cite{GS} for further details. 
In the next section we are going to discuss the implications of these 
parametrizations for the measurements of the double photon cross section 
at RHIC.

\section{Numerical Results}

We now turn to the numerical results for polarized isolated double photon
production at RHIC. It is still not clear what isolation parameters will
be used in the experiments, but preliminary studies \cite{yokosawa} seem
to indicate that an energy resolution parameter of $\epsilon = 0.5
\;{\rm GeV}/p_T^{\gamma}$ will be possible. We assume a cone size of $R=0.5$,
where $R$ is defined as the radius of a cone in the
pseudorapidity-azimuthal angle plane via the relation
\begin{equation}
R=\sqrt{(\Delta y)^2+(\Delta \phi)^2}.
\end{equation}
If a parton with energy fraction greater than $\epsilon$ times that of
the photon falls into the cone, the event is rejected.
 
As discussed in section II, we make use of the recent polarized proton
distributions evolved in NLO by Gehrmann and Stirling \cite {GS} and 
Gl\"{u}ck et al \cite{GRSV}. For the unpolarized proton distributions, 
we use the CTEQ3M set \cite{CTEQ} after checking that the GRV \cite{GRV}
distributions give very similar results. We also use fragmentation
functions for the photon evolved in NLO from ref.\cite{GRVP}. The value 
of $\Lambda_{QCD}$
used is chosen to correspond with the parton distribution set used. For
the electromagnetic coupling constant we use $\alpha_{em}=1/137$ and we
use the two-loop expression for $\alpha_s$. Unless otherwise stated we
set all renormalization/factorization scales to
$\mu^2=((p_{T1}^{\gamma})^2+(p_{T2}^{\gamma})^2)/2$, where
$p_{T1}^{\gamma}$ and $p_{T2}^{\gamma}$ are the transverse momenta of the 
first and second photons respectively. 

In Figs.1a and 1b we compare the different parametrizations of $\Delta
G(x,Q^2)$ as a function of $x$ at $Q^2=100$ GeV$^2$. The two GRSV gluons
are practically identical so we would not expect very different
predictions for any cross section sensitive to $\Delta G$ from these two
parametrizations. The three G-S distributions on the other hand, show 
much larger differences, and are thus likely to give different
predictions for processes depending on $\Delta G$. 

At the moment, it is proposed that RHIC will run at various
centre-of-mass energies between $\sqrt{S}=50$ to $500$ GeV, thus we
perform calculations for two values,  $\sqrt{S}=200$ and $500$ GeV.
It turns out that the NLO parametrizations for the polarized
distributions do not include charm quark distributions, thus in our
study we neglect the contribution for charm quarks in both the polarized
and unpolarized cases. It is expected that at RHIC cms energies,
the charm contribution will not be very large. 

In keeping with the convention started by the WA-70 collaboration
\cite{WA70}, and
which has been used in all subsequent analyses, we place asymmetric cuts on
the two photon $p_T's$. We require that the photon with the highest
transverse momentum has $p_T\geq 10$ GeV while the other is required to
have $p_T\geq 9$ GeV. This asymmetric cut ensures that the two-body
contributions to the cross section, such as the Born contribution, are not
favored over the three-body contributions. This differs from the way we
calculated the cross section in ref.\cite{corgor}, where we looked at
the $p_T$ distribution of one of the photons while placing cuts on the
other. In the present way of calculating the cross section, each photon
which passes the cut contributes to the cross section. This means we
have two entries for each event.  Latest indications are that the
maximum rapidity coverage achievable at RHIC will be in the range $-2 \leq
y \leq 2$, we thus restrict ourselves to this range. It is also
suggested that they will not be able to reliably detect photons with
$p_T\leq 10$ GeV.

For comparison, in Fig.2a we show the $p_T$ distribution for the
unpolarized case as predicted using the GRV and CTEQ3M parton
distributions at $\sqrt{S}=200$ GeV. They clearly give very similar 
results in the region tested, hence, in all subsequent discussions we will 
use only the CTEQ3M distributions. In Figs.2b and 2c we display the 
corresponding polarized cross sections. The results indicate that it
will be difficult to measure this cross section beyond about
$p_T=20$ GeV, even with the planned high luminosities at RHIC. More
interestingly, Fig.2b suggests,as we expected, that GRSV distributions give 
very similar
results and cannot be separated by this process. Fig2c suggests, on the
other hand, that the three G-S distributions give substantially different
predictions, and will probably be distinguishable. The G-S $a$
parametrization gives similar results to the two GRSV versions, as may
be expected from Fig1a. 

In Figs.3a and 3b we show the longitudinal asymmetries as predicted by the
various parametrizations. The asymmetry, $A_{LL}$, is defined by the
relation
\begin{equation}
A_{LL}=\frac{ \frac{d\Delta\sigma}{dp_T} }{\frac{d\sigma}{dp_T} },
\end{equation}
the ratio of the polarized to the unpolarized cross section, and gives a
measure of the spin dependence or spin sensitivity of the process. 
Again the two GRSV distributions, as expected, give
similar results, while the G-S ones give clearly distinguishable
results. The predictions also indicate that the asymmetries are not very
large in the measurable region of the cross section, varying between $5$
and $8\%$ for the GRSV and $4$ and $6\%$ for the G-S $a$
parametrization. It will require very high statistics measurements to
measure this asymmetry, but this may be achievable at RHIC, as long as
the cross section can be measured, given the planned detectors. 

At $\sqrt{S}=500$ GeV the situation improves somewhat with regards to the
size of the cross section. Fig.4a compares the $p_T$ distributions at
$\sqrt{S}=500$ and $200$ GeV for the unpolarized cross section. As
expected, at higher cms energies the cross section is substantially
larger. In Fig.4b we show the polarized cross section at
$\sqrt{S}=500$ GeV. On the figure we also show the full prediction given by
the G-S $a$ distribution as well as the contribution form $q g$
initiated subprocesses. In the case of G-S $a$ the contribution for the
$q\bar{q}$ initiated subprocess (not shown) is negative. Corresponding
distributions are shown for the GRSV 1 parametrization. In this case the
$q\bar{q}$ process gives a positive contribution but it is substantially
smaller than the $q g$ one. The G-S $a$ and GRSV 1 distributions predict
similar cross sections, but the relative importance of the subprocesses
is clearly very different. The most interesting aspect of this result
from the point of view of sensitivity to $\Delta G$ is the fact that in
both cases, the $q g$ initiated process dominates. We should mention
that in this calculation we do not include contributions for the higher
order process ($O(\alpha_s^2)$) $gg\rightarrow \gamma\gamma$, preferring to keep
consistently to $O(\alpha_s)$. 

Figs.4c and 4d show the asymmetries as predicted by the G-S $a$ and GRSV
polarized distributions. Included also are the contributions to the
asymmetries from the $qg$ and $q\bar{q}$ initiated processes. In Fig.4d
we also include the asymmetry in LO where we have used the LO
counterpart of the GRSV 1 parametrization. The LO cross section consists
only of the process $q\bar{q}\rightarrow \gamma \gamma$ plus the
fragmentation processes
$q\bar{q}\rightarrow g \gamma$ plus $q g\rightarrow q \gamma$. It can be
said that the asymmetry is fairly stable under the higher order
corrections, but the corrections are clearly significant. As we expected 
from the results shown in Fig.4b, the
NLO asymmetry is dominated by the $qg$ initiated process. this is the
case when either distributions is used. On the other hand, the size of
the asymmetry is not very different from that at $\sqrt{S}=200$ GeV,
varying between $3$ and $8\%$ over the measurable range. 
 
As an indication of the stability of our predictions we show the scale
dependence of the unpolarized cross section in Fig.5. The cuts and
distributions are the same as for the solid curve in Fig.4a. By varying
all renormalization and factorization scales in the range $1/2\leq n
\leq 2$, where $\mu=n( (p_{T1}^\gamma)^2+(p_{T2}^\gamma)^2)/2$, we can
vary the cross section by as much as $20\%$ in the region $10 \leq
p_T\leq 35$. This is still a substantial uncertainty in the predictions,
but it is a significant reduction over the corresponding figure of
$60\%$ for the LO predictions. 

Finally in Fig.6 we show the K-factor as a function of $p_T$ where,
\begin{equation}
K=\frac{   \frac{d\sigma(LO)}{dp_T^\gamma}   }{
\frac{d\sigma(NLO)}{dp_T^\gamma} },
\end{equation}
at $\sqrt{S}=500$ GeV, for the unpolarized cross section. 
This confirms that the NLO corrections are indeed quite significant at the
$p_T$ values relevant at RHIC. 

\section{Conclusions}

An $O(\alpha_s)$ NLO calculation of the process $pp\rightarrow \gamma
\gamma+X$ was presented, where for the first time polarized parton 
distributions evolved in NLO QCD is used. The two photons were also
isolated using plausible isolation parameters. We found that the cross
section for the precess will not be very large at RHIC, particularly at
the lower CMS energies, but it will nevertheless still be measurable in
the lower $p_T^\gamma$ region. The new polarized parton
distributions predict an asymmetry of between $3$ and $8\%$ in the
measurable region for the process. While the asymmetry cannot be said to
be large, given sufficiently good statistics, it should still me
measurable. Given this possibility, we found that a discrimination
between extreme parametrizations of the polarized gluon distribution
$\Delta G$ should be possible. This is due in particular to the
dominance of the subprocess $q g\rightarrow\gamma \gamma q$ over the
$q\bar{q}$ annihilation process. We thus conclude that this process, if
studied at RHIC could prove useful for supplementing information on the
polarized distributions obtained form other sources.

\section{Acknowledgements}
We would like to thank Bob Bailey for help with the programing aspects
of the Monte Carlo method used in this calculation. We thank G. Ramsey, 
A. Yokosawa for discussions. C.C. is grateful to Sang Hyeon Chang for 
helpful discussions. 
This work supported in part by the U.S.~Department of Energy, Division 
of High Energy Physics, Contract W-31-109-ENG-38 and DEFG05-86-ER-40272.

\pagebreak


\pagebreak

\noindent
{\bf Figure Captions}

\newcounter{num}
\begin{list}%
{[\arabic{num}]}{\usecounter{num}
    \setlength{\rightmargin}{\leftmargin}}

\item (a) The polarized gluon distribution in NLO as a function of $x$ at
$Q^2=100$ GeV$^2$ for the G-S $a$ (solid line) GRSV1 (dashed line) and
the GRSV2 (dotted line). (b) Same as (a) but for the G-S $a$, G-S $b$
and G-S $c$ parametrizations.

\item (a) The unpolarized cross section $d\sigma/dp^\gamma_T$ for
$-2\leq y^\gamma \leq 2$ as a function for $p_T^\gamma$ at $\sqrt{S}=200$ GeV as
predicted by the GRV (solid line) and CTEQ3M (dashed line) parametrizations 
of the proton distribution functions. (b) Same as (a) but for the
polarized cross section as given by the GRSV1 (solid line) and GRSV2
(dashed line) parametrizations of the polarized parton distributions.
(c) Same as (b) but for the G-S $a$ (solid line) G-S $b$ (dashed line) and
G-S $c$ (dot dashed) line. The GRSV1 (dotted line) is included for
comparison. 

\item (a) The longitudinal asymmetries for the cross sections displayed in 
Fig.2b. (b) Same as (a) but for the G-S polarized distributions. The
unpolarized distributions used is CTEQ3M. 

\item (a) Comparison of the unpolarized cross section $d\sigma/dp^\gamma_T$
as a function of $p_T^\gamma$ for $-2\leq y^\gamma\leq 2$ using the
CTEQ3M proton distributions, at $\sqrt{S}=200$ (dashed line) and
$\sqrt{S}=500$ GeV (solid line). (b) The polarized cross section at 
$\sqrt{S}=500$ GeV
using the G-S $a$ (solid line) and GRSV1 (dot dashed line). Included for
comparison are the contributions from the $qg$ initiated process. In the
case of G-S $a$ it is the dashed line and for GRSV1 the dotted line.
(c) The asymmetry predicted by the G-S $a$ parametrization for the cross
section given in (b). Included for comparison are the asymmetries as
given by the $qg$ (dashed line) and $q\bar{q}$ (dotted line) initiated
processes. (d) Same as (c) but for the GRSV1 distributions. The LO
prediction for the asymmetry (dot dashed line) using the LO version of
the GRSV1 distribution is included. 

\item The renormalization/factorization scale ($\mu$) dependence
for the unpolarized cross section given in Fig.4a, for three different 
choices of scale,  $\mu=n( (p^\gamma_{T1})^2+(p^{\gamma}_{T2})^2)/2$: 
0.5, 1.0, and 2.

\item The K-factor for the unpolarized cross section given in Fig.4a at
$\sqrt{S}=500$ GeV, where
K=$d\sigma(LO)/dp_T^\gamma/d\sigma(NLO)/dp_T^\gamma$. 

\end{list}

\epsffile{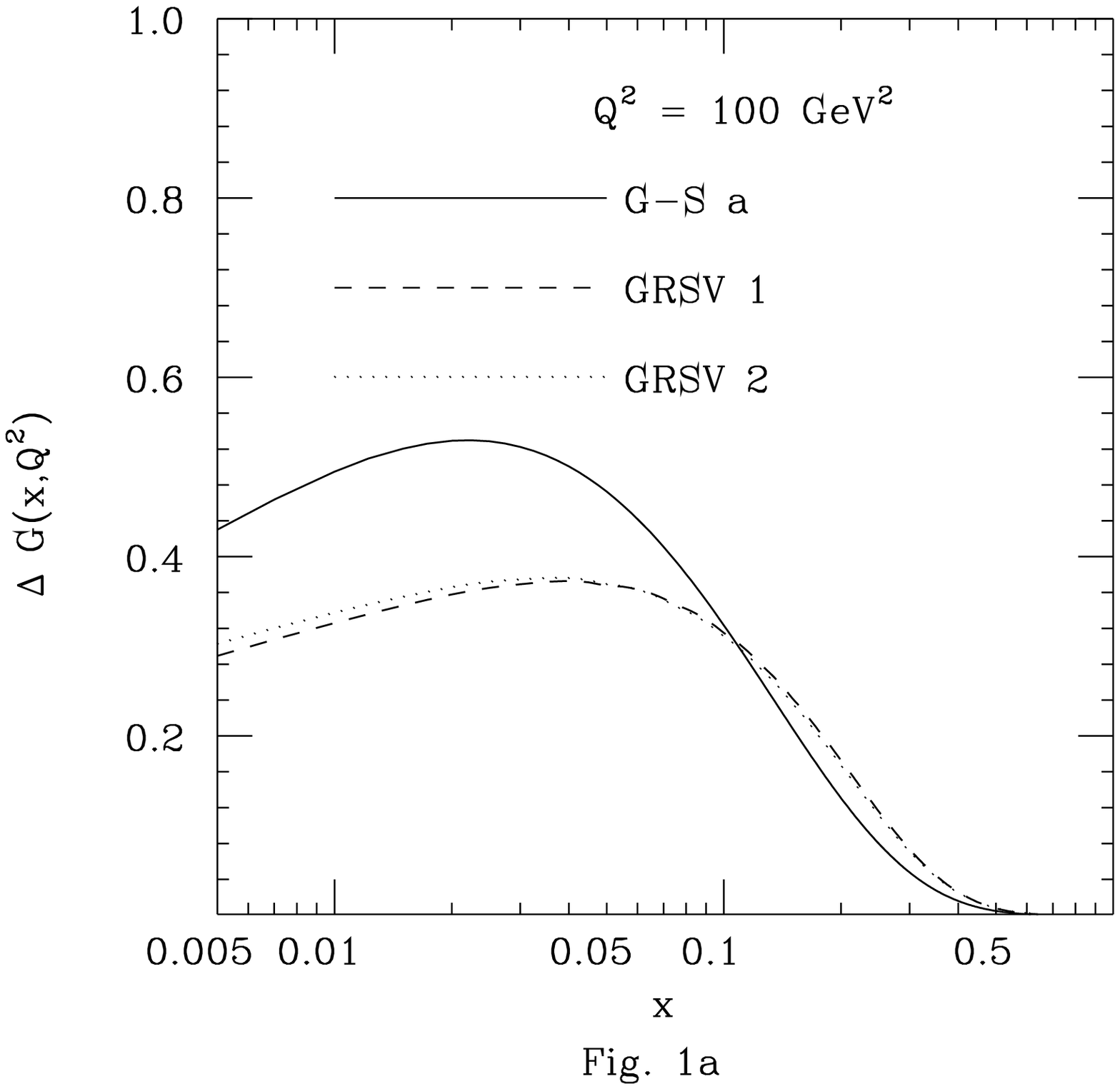}
\epsffile{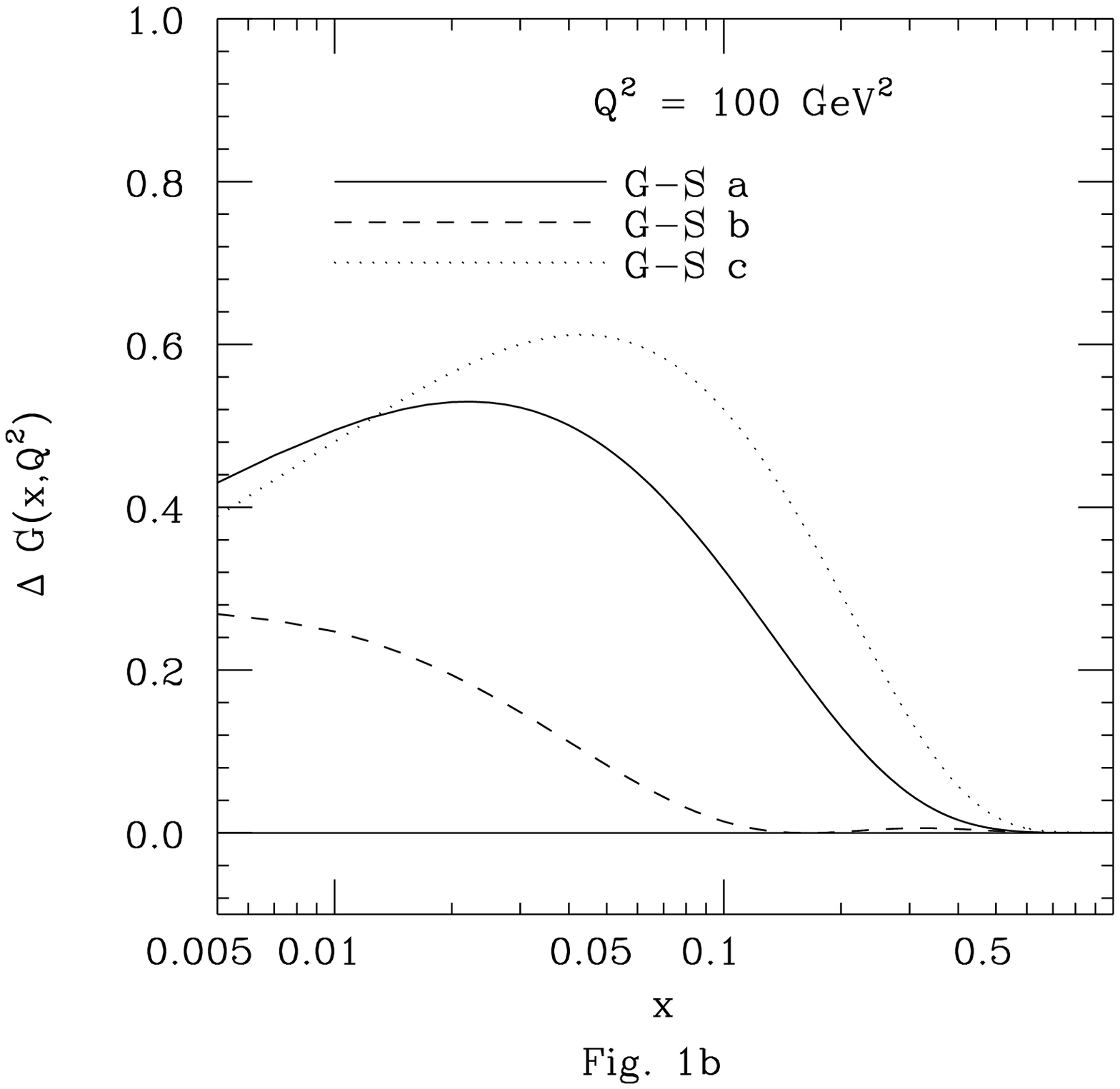}
\epsffile{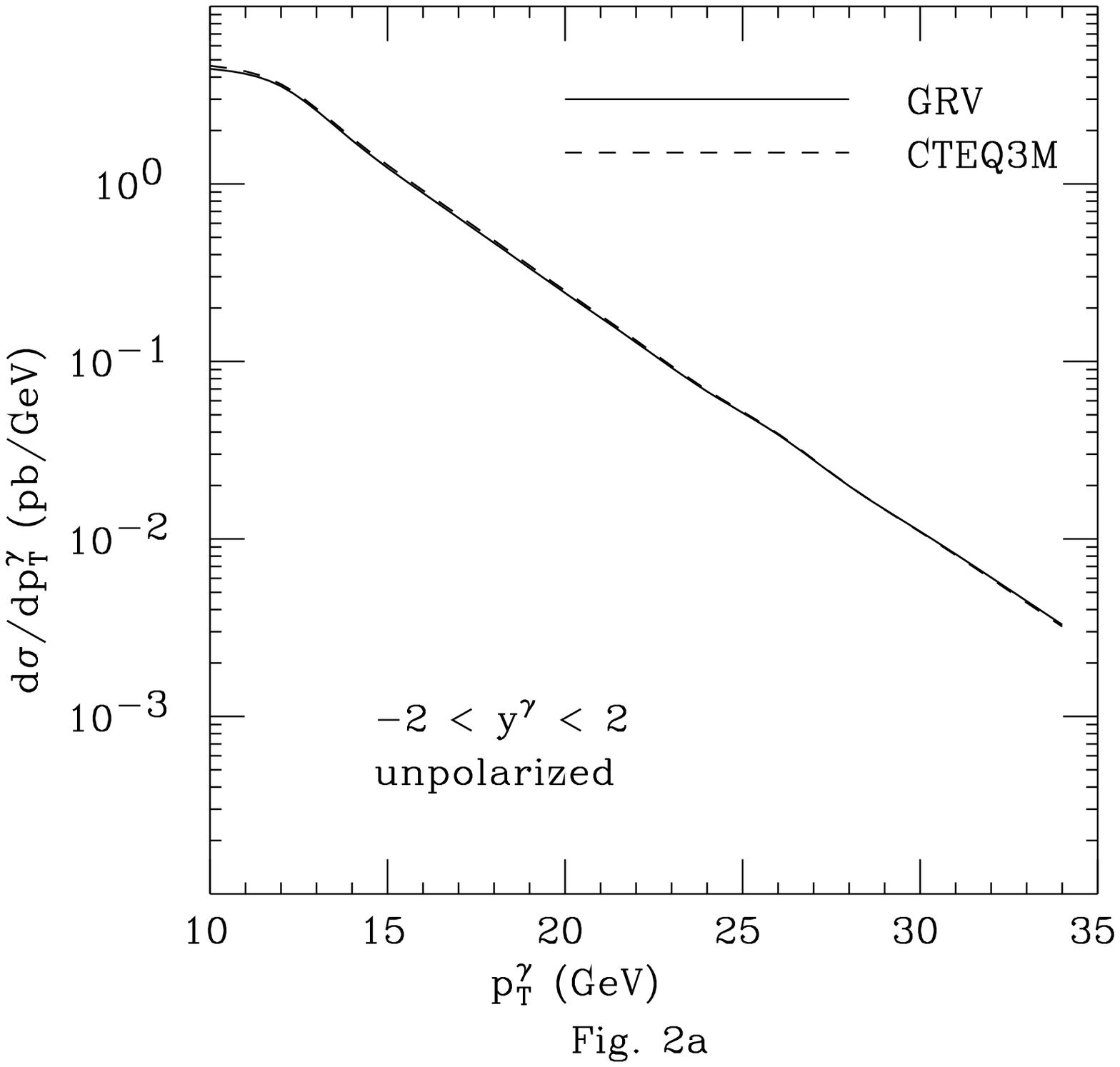}
\epsffile{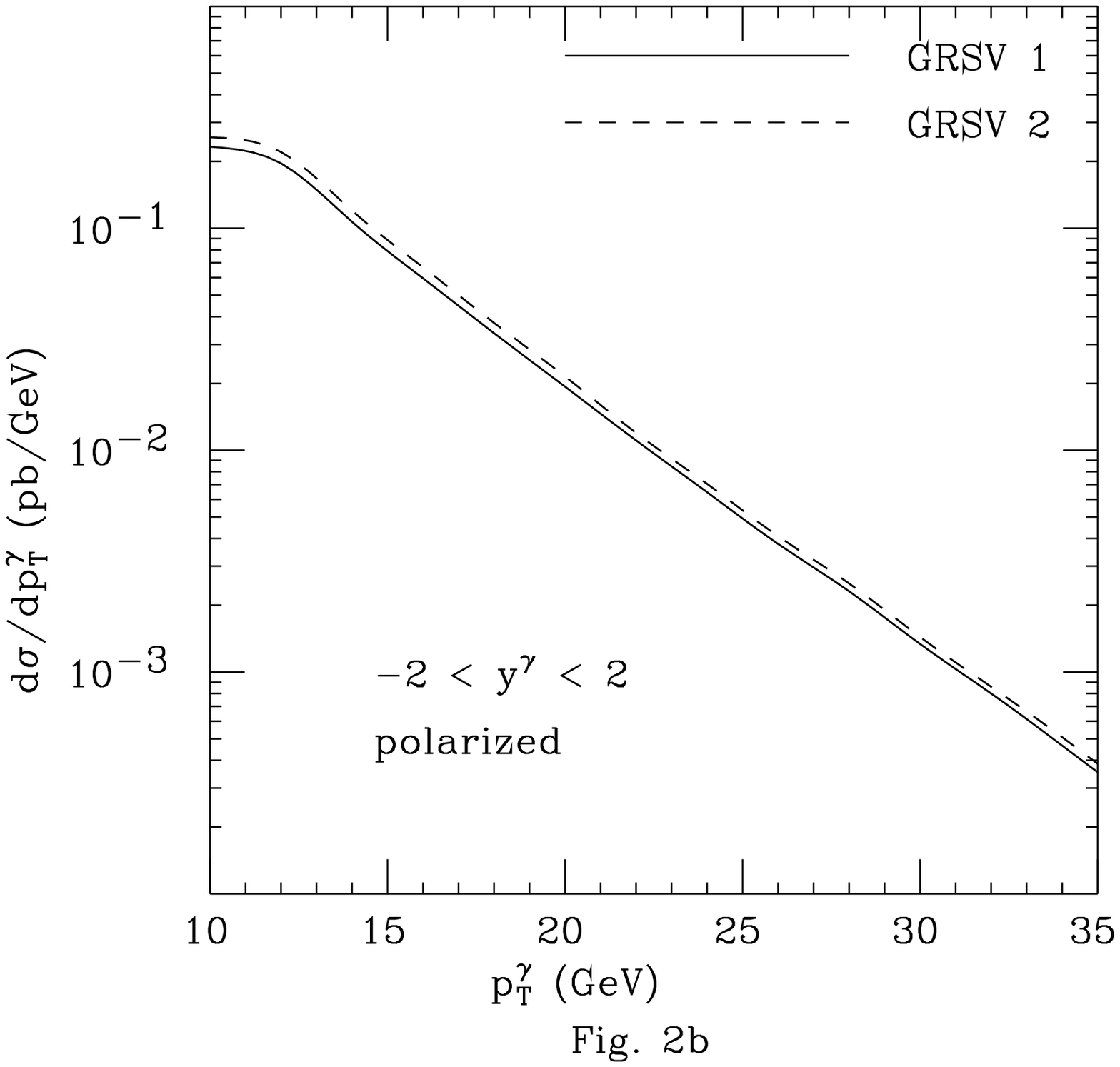}
\epsffile{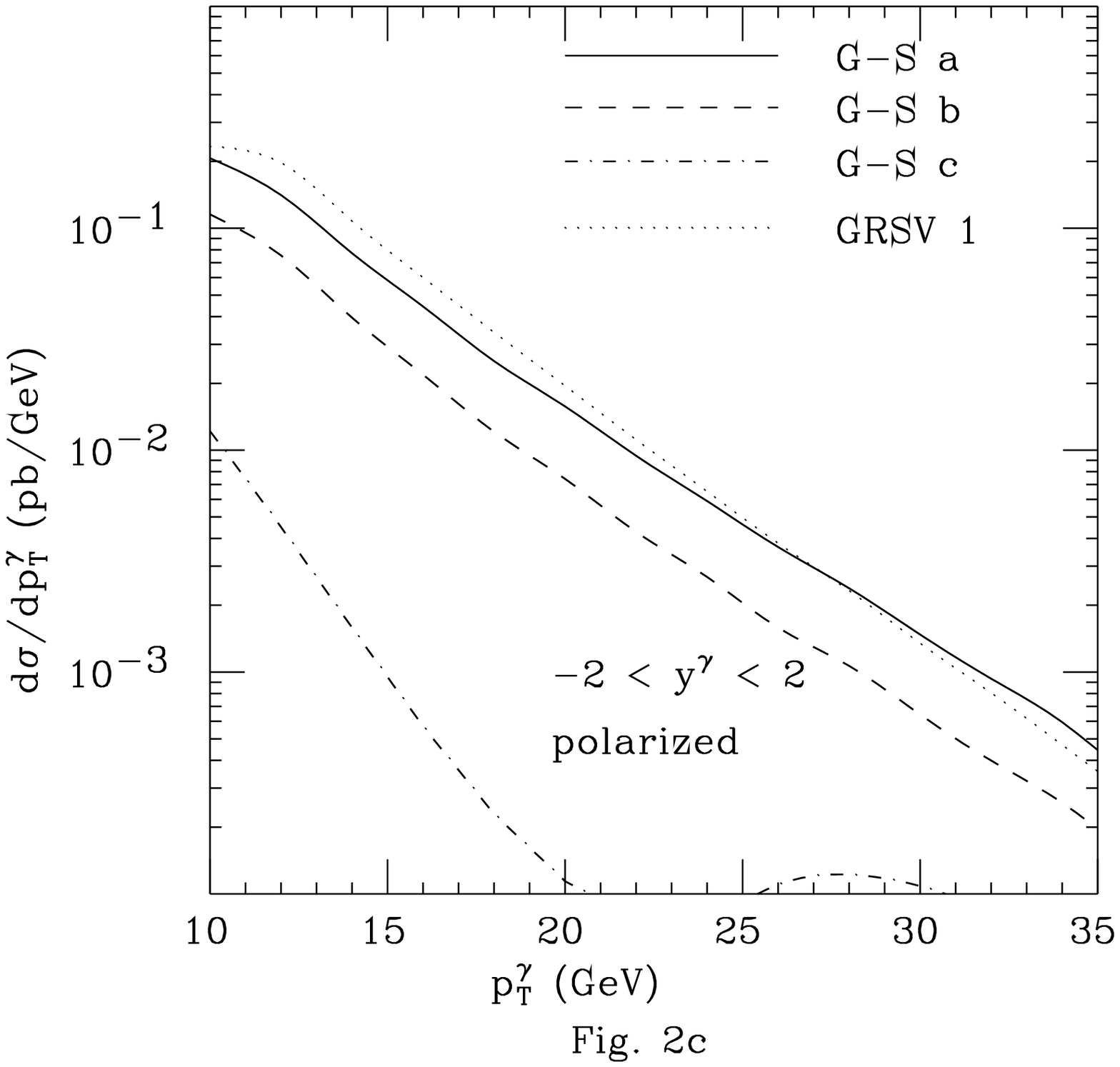}
\epsffile{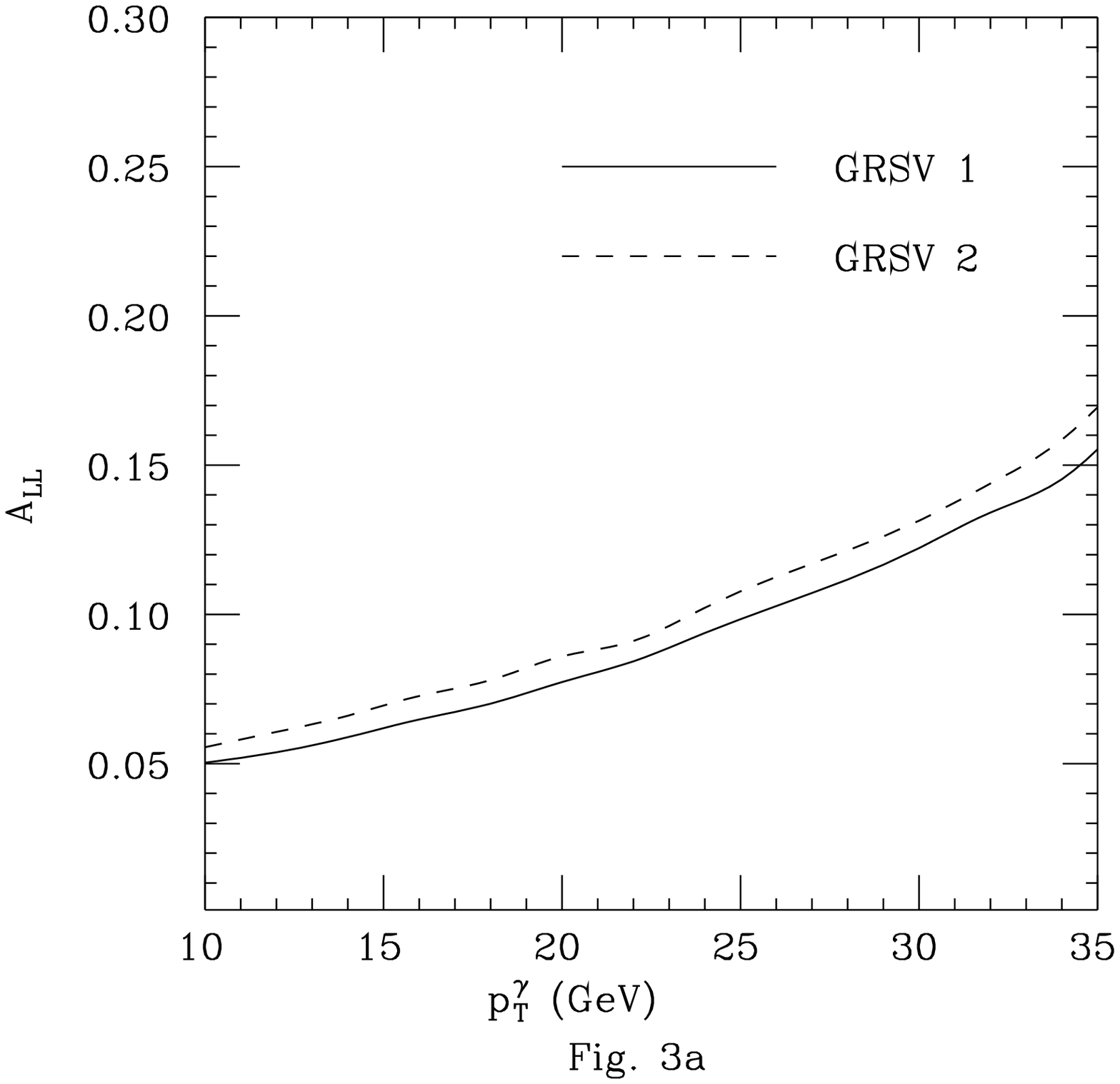}
\epsffile{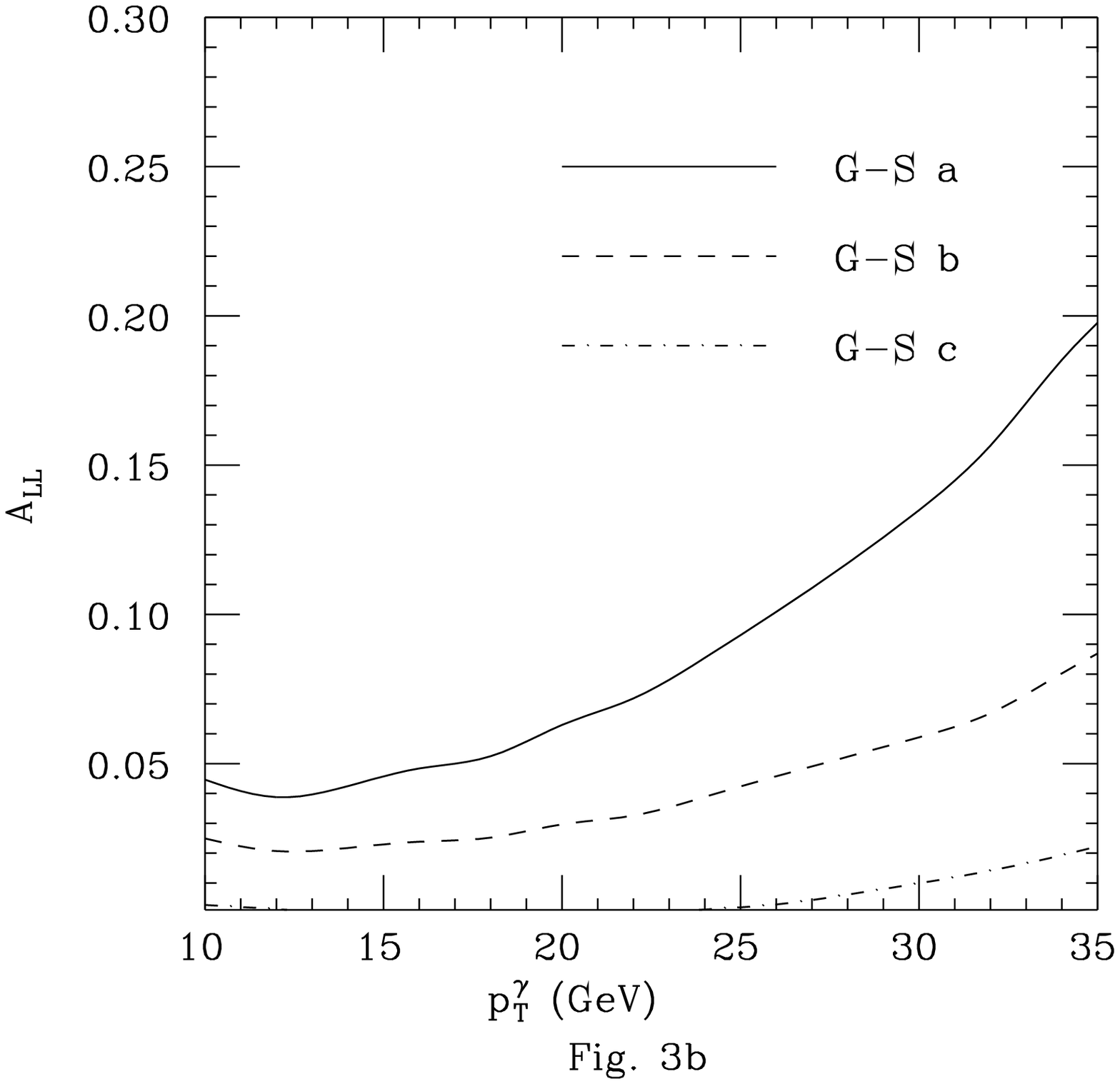}
\epsffile{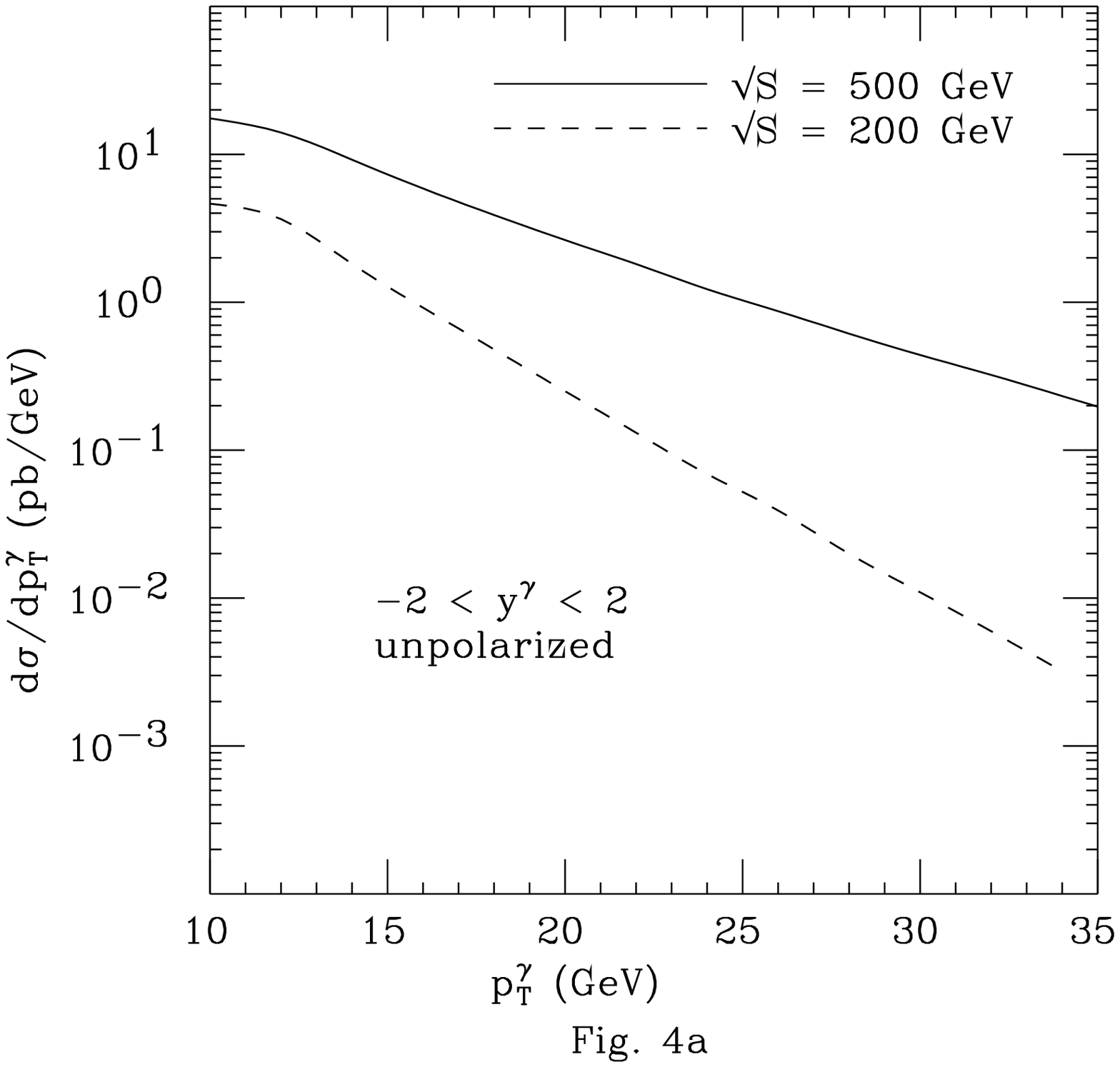}
\epsffile{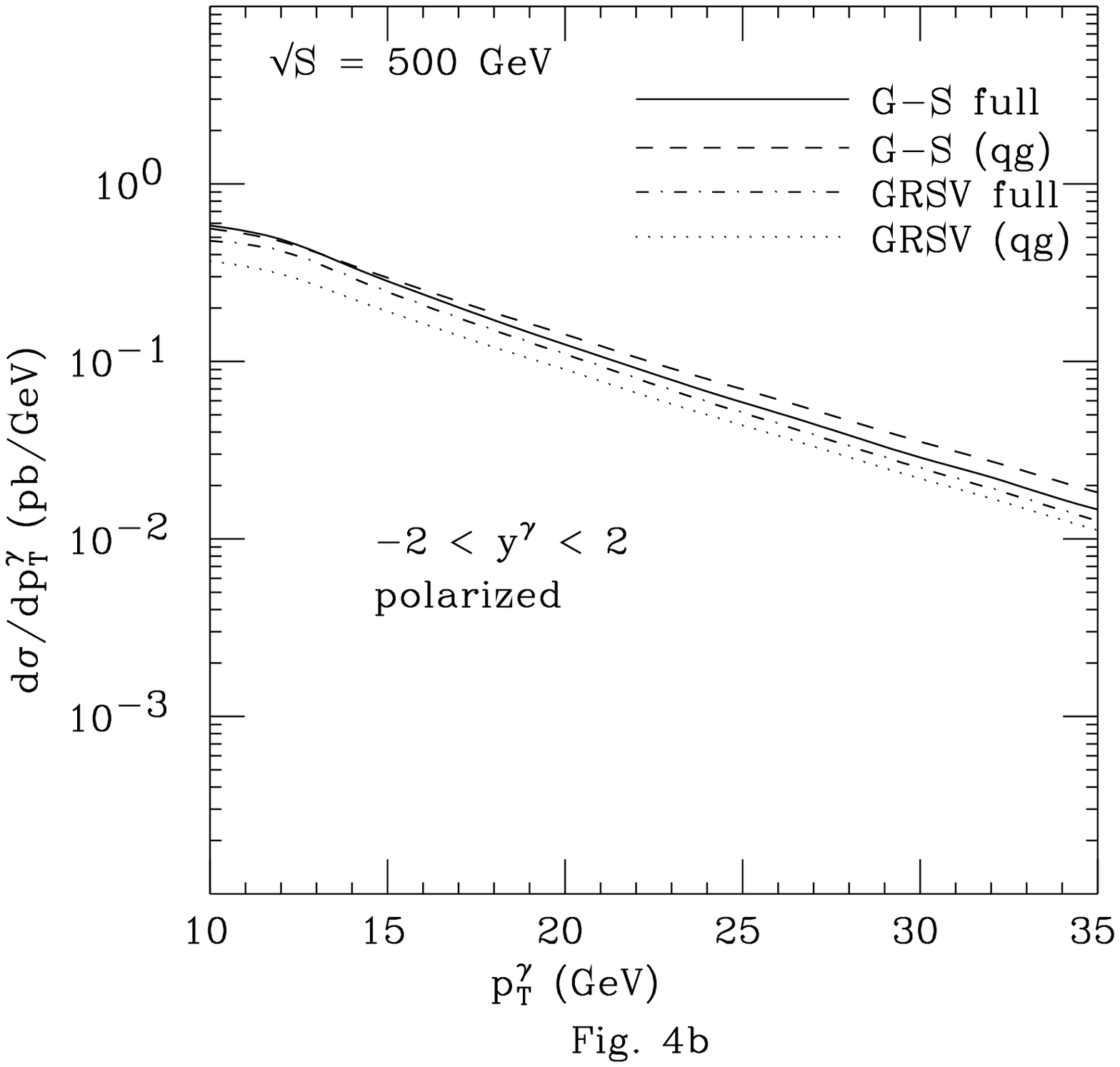}
\epsffile{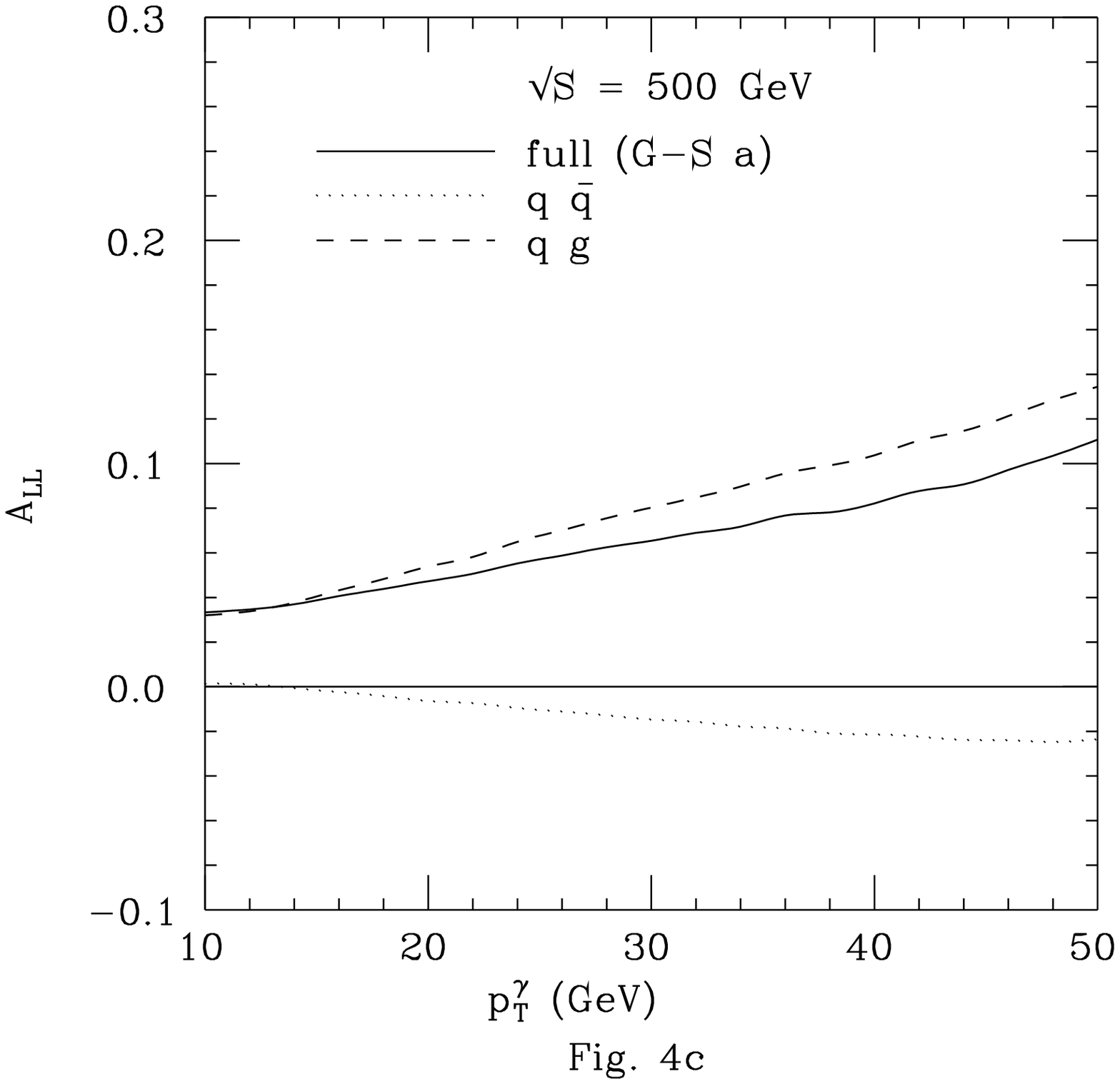}
\epsffile{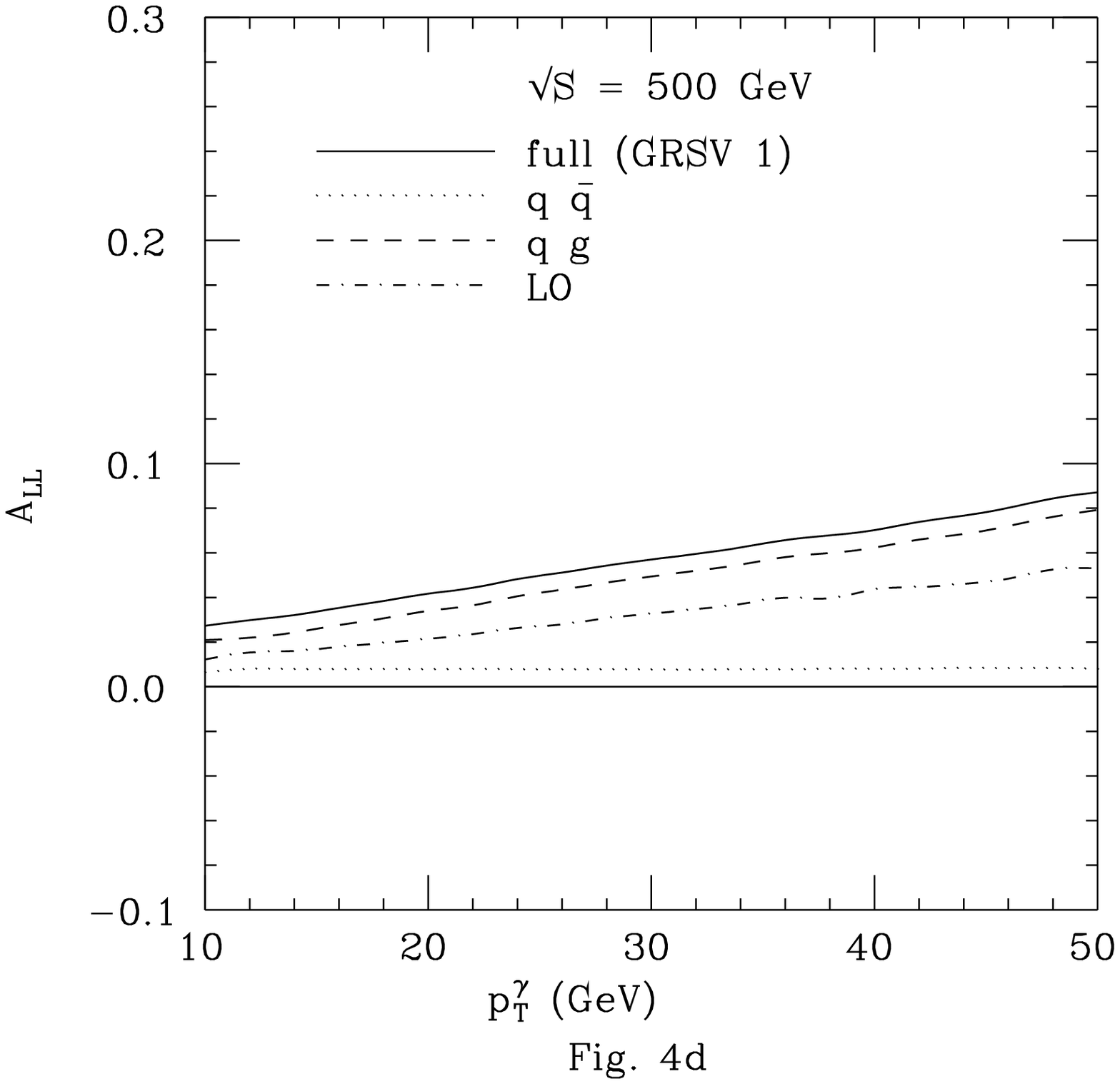}
\epsffile{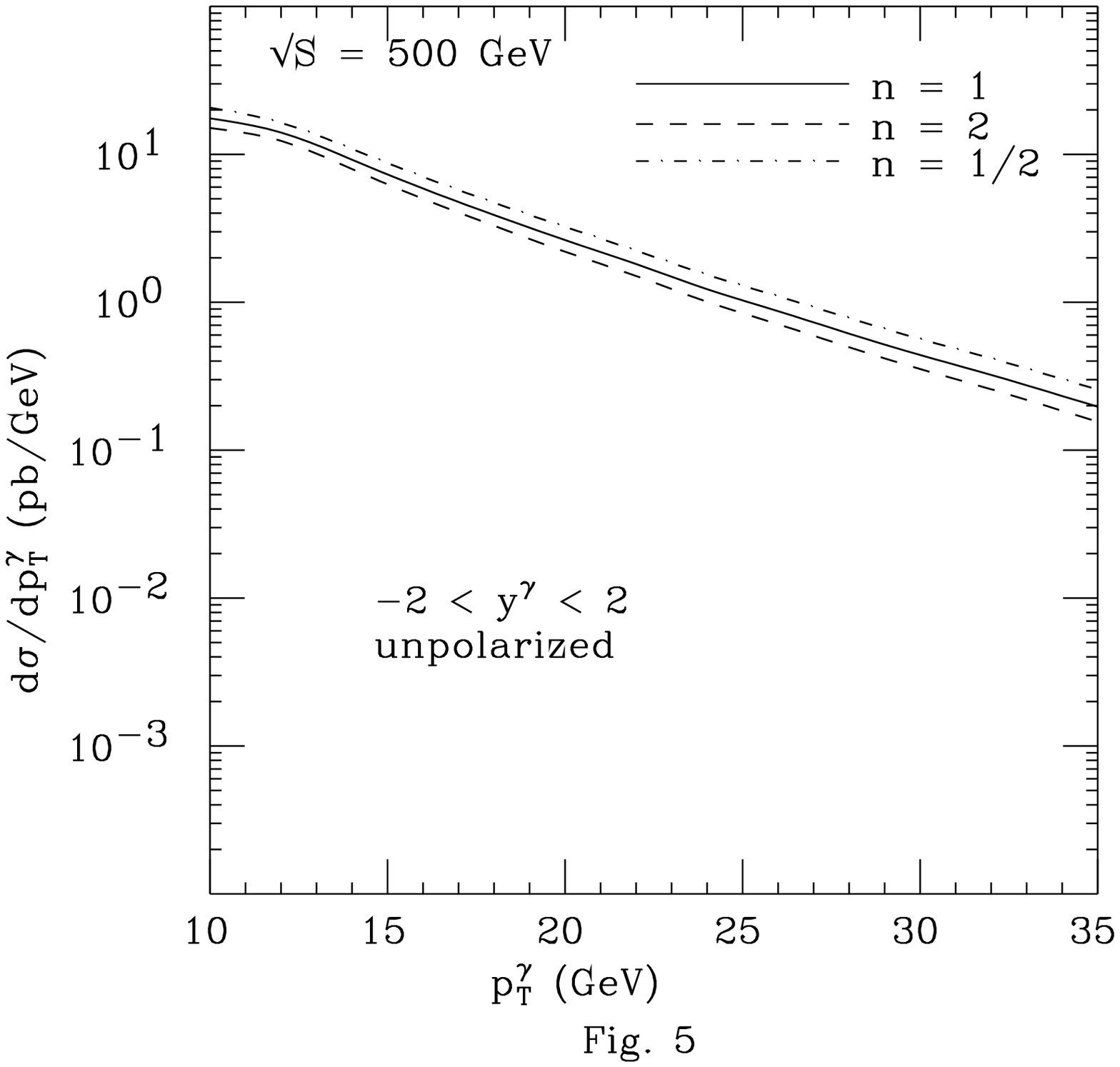}
\epsffile{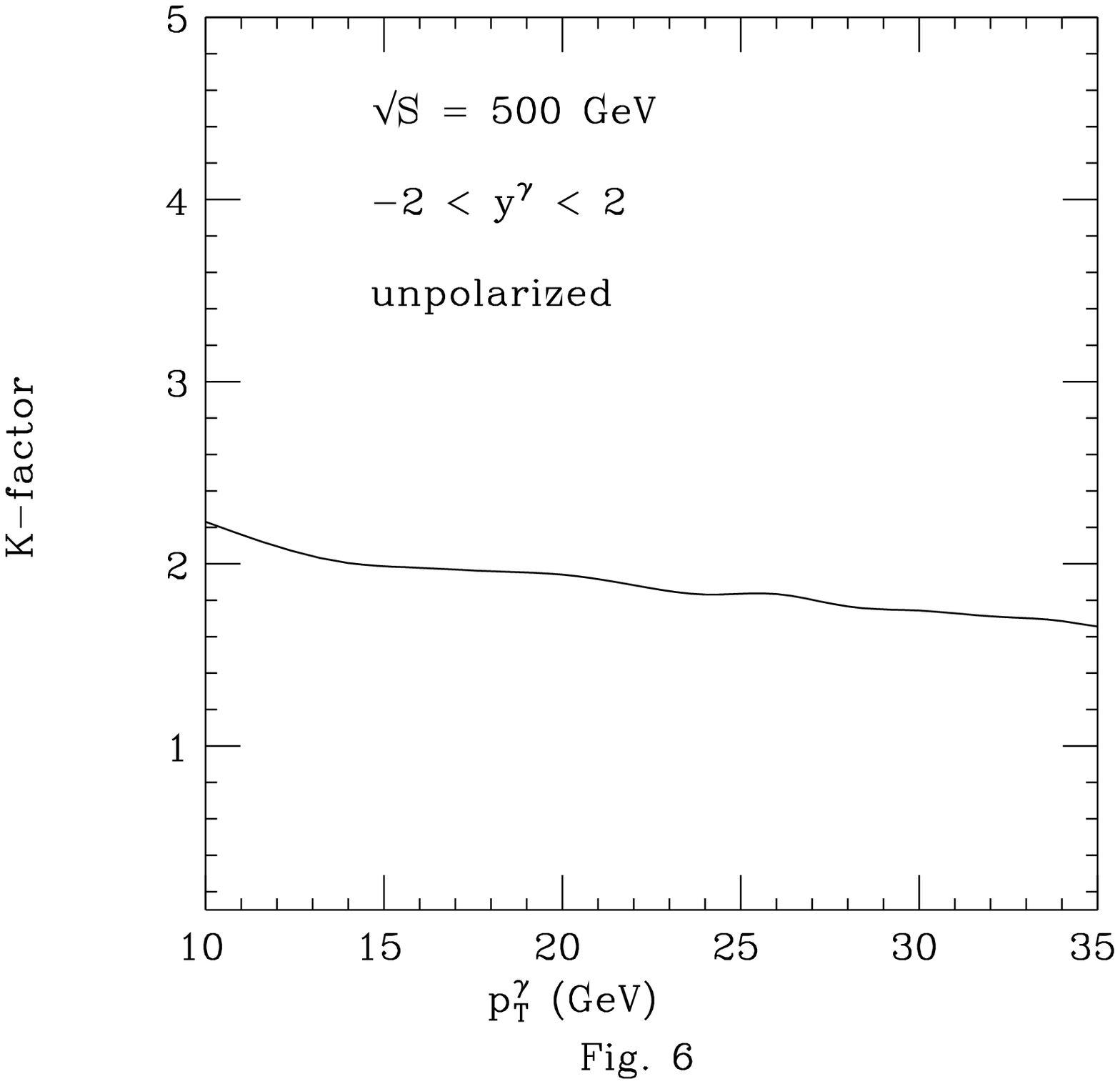}


\begin{thebibliography}{99}
\bibitem{EMC} J. Ashman et al., EMC collaboration, Phys. Lett. B206 (1988) 364, 
Nucl. Phys. B328 (1989) 1.
\bibitem{set2} SMC Collaboration, B Adeva et al. Phys. Let. B302 (1993) 533, 
ibidem B320 (1994)400;
 SMC Collaboration, D. Adams et al., Phys. Lett. B329 (1994) 399;
E143 Collaboration, K. Abe et al., SLAC-PUB-6508 (1994) preprint;
E143 Collaboration, R. Arnold et al., presented at ICHEP94, 
Glasgow, August 1994.  
\bibitem{WA2} WA2 Collaboration, Z. Phys. {\bf C 21} (1983), 27; Particle Data 
Group, Phys. Lett. {\bf B111} 1982. 
\bibitem{HSU} S. Y. Hsu et al., Phys. Rev. {\bf D 38} 2056 (1988). 
\bibitem{MVN}R. Mertig and W. L. Van Neerven 
NIKHEF-H/95-031. W. Vogelsang, Rutherford Preprint, RAL-TR-95-071.
\bibitem{yokosawa} A. Yokosawa, Private Communication.
\bibitem{GS} T. Gehrmann and W. J. Stirling, Durham Preprint DTP/95/82.
\bibitem{GRSV} M. Gl\"{u}ck, E. Reya, M. Stratmann and W. Vogelsang,
Dortmund Preprint DO-TH 95/13 and Rutherford Preprint RAL-TR-95-042.
\bibitem{AP} G. Altarelli and G. Parisi, Nucl. Phys. {\bf B 126} (1977) 298.
\bibitem{CTEQ} H. L. Lai {\it et al.,} CTEQ Collaboration, Phys. Rev.
{\bf D51}, 4763 (1995).
\bibitem{GRV} M. Gl\"{u}ck, E. Reya and A. Vogt, Z. Phys. {\bf C 67} 433 (1995).
\bibitem{GRVP} M. Gl\"{u}ck, E. Reya and A. Vogt, Phys. Rev. {\bf D48},
116 (1993).
\bibitem{Lip} H.J. Lipkin, Phys. Lett. {\bf B 256}, 284 (1991); 
{\bf B 337}, 157 (1994).
\bibitem{Ji} X. Ji, MIT-CTP-2411 hep-ph/{\bf 9502288}.
\bibitem{bd} S. J. Brodsky and I. Schmidt, Phys. Lett. {\bf B 234}, 
144, (1990); S. J. Brodsky, M. Burkhardt and I. Schmidt, Nucl. Phys. 
{\bf B441}, 197 (1990).
\bibitem{WA70} E. Bovin {\it et al.,} Z. Phys. {\bf C41}, 591 (1989).
\bibitem{corgor} C. Corian\`{o} and L. E. Gordon Argonne Preprint,
ANL-HEP-PR-95-84 and IFT-UFL-95-28.
\end{thebibliography}
\end{document}